\documentclass[aps,prl,twocolumn,floats,showpacs]{revtex4}
\usepackage{graphicx}
\usepackage{amsmath}
\usepackage{amssymb}

\newcommand{\1}{{|1\rangle}}
\newcommand{\2}{{|2\rangle}}
\newcommand{\A}{\bar{\alpha }}

\renewcommand{\ap}{{\alpha_P}}
\newcommand{\af}{{\alpha_F}}
\newcommand{\R}{{\rm R}}

\begin{document}
\title{Population transfer between two quantum states by piecewise chirping of femtosecond pulses: Theory and experiment}

\date{\today}
\author{S.~Zhdanovich$^{1,3}$, E.A.~Shapiro$^2$, M. Shapiro$^{1,2,3}$, J.W.~Hepburn$^{1,2,3}$, and V.~ Milner$^{1,2,3}$}
\affiliation{Departments of  Physics \& Astronomy$^1$ and Chemistry$^2$, and The Laboratory for Advanced Spectroscopy and Imaging Research (LASIR)$^3$, The University of British Columbia, Vancouver, Canada}

\begin{abstract}{We propose and demonstrate the method of population transfer by piecewise adiabatic passage between two quantum states. Coherent excitation of a two-level system with a train of ultrashort laser pulses is shown to reproduce the effect of an adiabatic passage, conventionally achieved with a single frequency-chirped pulse. By properly adjusting the amplitudes and phases of the pulses in the excitation pulse train, we achieve complete and robust population transfer to the target state. The effect is demonstrated experimentally by observing piecewise excitation of Rubidium atoms from $5s_{1/2}$ to $5p_{1/2}$ electronic state. We show that similarly to the conventional adiabatic passage, the piecewise process is insensitive to the total excitation energy as long as the adiabaticity conditions are satisfied. The piecewise nature of the process suggests that robust and selective population transfer could be implemented in a variety of complex quantum systems beyond the two-level approximation.}
\end{abstract}


\maketitle

The existence of robust and selective methods of executing population transfer between quantum states is essential for a variety of fields, such as, precision spectroscopy and atomic clocks\cite{Diddams04}, quantum computing\cite{NielsenChuangBook}, control of molecular dynamics and chemical reactions\cite{ShapiroBrumerBook}. Traditionally, population transfer between two quantum states has been achieved by either executing a half-cycle Rabi flopping, i.e. by applying a ``$\pi $-pulse'', or by optically inducing an adiabatic passage (AP) between the states of interest\cite{ShoreBook,AllenEberlyBook}. Though the application of a $\pi$-pulse can be implemented on a very short time scale, it is far from being robust, as it is highly sensitive to fluctuations in the laser power, phase, and pulse duration. In contrast, AP, which for example can be executed by the STIRAP technique\cite{Gaubatz90}, or by ``slowly'' chirping the instantaneous frequency of the pulse\cite{Vitanov01}, exhibits high degree of robustness against the fluctuations of many of the laser field parameters. Because of this property, AP with chirped pulses has been widely used for executing population transfers in atomic\cite{Broers92,Chatel03} and molecular\cite{Chelkowski90,Melinger94} systems, including recent applications to enhancing the photo-association of ultracold molecules\cite{Brown06,Wright07}.

Though well suited for two-level systems, population transfer with chirped pulses in multilevel systems becomes sensitive to the exact value of the chirp and field strength\cite{Cao98,Malinovsky01}, thus losing some of its appeal as a robust way of efficiently manipulating populations. When used with spectrally broad ultrashort pulses, pulse chirping can no longer selectively populate a pre-chosen single state or a superposition of states\cite{Melinger94,Malinovsky01}, and other methods of adiabatic\cite{Kral07} or non-adiabatic\cite{Dudovich02,Hauer06} excitation must be employed. Since ultrashort pulses do not satisfy the adiabatic conditions, it is customary to treat the problem in a purely empirical way by designing feedback-controlled experiments with genetic search algorithms\cite{Judson92,Prokhorenko05}. However, these approaches lack the robustness and efficiency of the AP method.

Recently, we have demonstrated theoretically that one can implement AP with ultra-short pulses by executing the transfer of population in a \emph{piecewise manner}\cite{Shapiro07, Shapiro07a}. The original work on Piecewise Adiabatic Passage (PAP)\cite{Shapiro07} proposed using two temporally overlapping pulse trains. It was shown that by tuning the parameters of the pulse trains one can bring about a complete population transfer between two quantum states through a third intermediate level. In this Letter we present the first experimental demonstration of the PAP method with a \emph{single} pulse train by introducing the technique of ``piecewise chirping''. We show that the piecewise population transfer faithfully reproduces an AP process executed with a continuous frequency-chirped pulse, achieving the same level of robustness and efficiency.

Consider a system of two states, $\1$ and $\2$, of energies $E_{1}$ and $E_{2}$, respectively. An arbitrary coherent superposition of these states can be written in the interaction representation in terms of two angles, $\theta $ and $\phi,$ as
\begin{equation}
\psi= \cos\theta/2 \exp[-i E_1 t]\, |1\rangle + \sin\theta/2
\exp[-i E_2 t + i\phi]\, |2\rangle~.
\label{twostate}
\end{equation}
This superposition state can be represented as a unit Bloch vector pointing along the $(\theta ,\phi )$ direction.
\begin{figure}
\centering
\includegraphics[width=0.98\columnwidth]{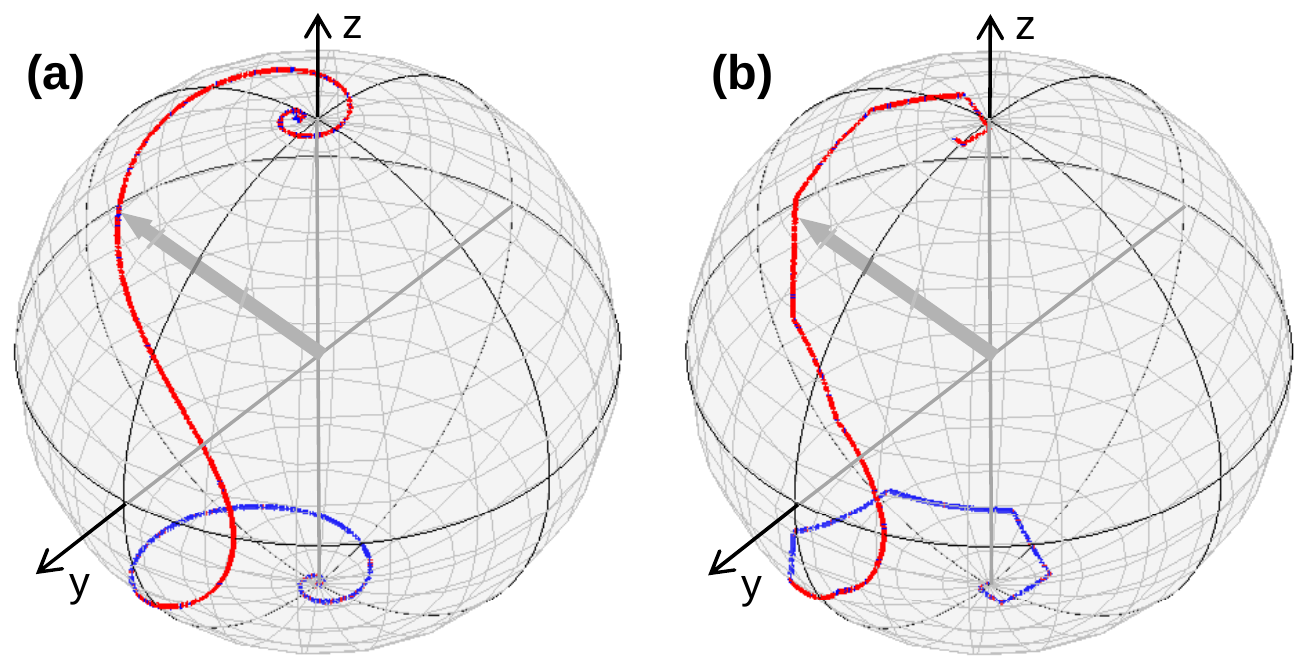}
\caption{(Color online). Two sample trajectories of the Bloch vector (thick gray arrow) during the AP process implemented with a single continuous chirped pulse \textbf{(a)}, and a train of 20 transform-limited pulses with quadratic temporal phase \textbf{(b)}(see text).}
  \vskip -.1truein
  \label{FigSpheres}
\end{figure}

In conventional AP with chirped pulses, the system is driven by a near-resonant field with a time dependent frequency detuning $\Delta $ relative to the exact resonance frequency $\omega _{0}=(E_2-E_1)/\hbar$. The product of the field strength and the transition dipole matrix element can be conveniently expressed in terms of a Rabi frequency $\Omega$. Application of the Rotating Wave Approximation in a frame rotating at the transition frequency $\omega _{0}$, results in two dressed eigenstates, $\psi_+$ and $\psi_-$. These dressed states constitute two stationary points on the Bloch sphere, $(\theta_\pm,\phi_{\pm})$, where $ \tan{\theta_\pm} =\pm\Omega/\Delta$, $\phi_\pm=\int\Delta(t) dt \pm \pi /2$. All other solutions of the time dependent Schr\"odinger equation precess around the axes defined by the above stationary points, displaying periodic trajectories corresponding to the familiar Rabi oscillations\cite{AllenEberlyBook}.

If the parameters of the field are varied adiabatically, and if initially $\Delta\ll-|\Omega|$, and at the end of the process $\Delta\gg|\Omega|$, there is an interchange of populations between state $|1\rangle$ and state $|2\rangle$. This interchange is expressed on the Bloch sphere as the movement of the Bloch vector from the $\theta = 0$ direction to the $\theta=\pi$ direction and  \emph{vice versa}. The motion proceeds via the spiral trajectory shown in Fig. \ref{FigSpheres}(a). The above behavior holds as long as the adiabaticity condition is maintained\cite{Vitanov01a}:
\begin{equation}\label{QuantumAdiabaticity}
|\dot\theta_M|\ll \sqrt{\Delta^2(t)+\Omega^2(t)}
\end{equation}

We now consider the evolution of the Bloch vector under the action of a train of short, mutually coherent, laser pulses. Each pulse, centered at the resonant frequency $\omega _{0}$ and of duration $\tau $, generates a rotation ${\hat P}\equiv\R_y(\ap)$ of the Bloch vector about the $y$ axis $(\pi /2,\pi /2)$ by an angle $\ap=\int_{\tau} \Omega (t) dt $. To account for the change in the carrier phase between two consecutive pulses we introduce an additional  $z$-rotation on the sphere ${\hat F}\equiv\R_z(\af)$. Thus, the overall evolution may be represented by a sequence of rotations $\hat U = ...\hat F \hat P \hat F \hat P ...$

The product $\hat F \hat P$ of two rotations of the Bloch vector is an overall rotation  by an angle $\alpha_0$ about an axis defined by the ($\theta_0,\phi_0$) angles, given to lowest-order expansion in $\ap$, $\af$ as
\begin{eqnarray}
\alpha_0      &=&  \sqrt{(\ap^2+\af^2)/2}, \label{a0}
\\
\phi_0        &=&  \pm\pi/2-\af/2,       \label{phi0}\\
 \tan\theta_0 &=&  \pm \ap/\af.          \label{theta0}
\end{eqnarray}
By maintaining the same value of $\af$ and $\ap$ throughout the pulse train we induce piecewise rotations of the Bloch vector around the closest stable point $(\theta_0,\phi_0)$. By slowly varying the values of $\ap,\af$ we can make the stable points move and the Bloch vector, captured near them, to follow faithfully. Intuitively, the conditions of such piecewise following are: (i) the $y$- and $z$-rotations should be small (i.e. that each pump pulse should induce an angular change much smaller than $\pi$ and each increment in the carrier phase should be small too), and (ii) ($\theta_0,\phi_0$) should move little during each pulse, i.e.
\begin{equation}\label{ClassicalAdiabaticity}
\Delta\theta_0 \ll \sqrt{(\ap^2+\af^2)/2}.
\end{equation}
If initially $\ap\ll |\af|$, the two stationary points, $\theta_{0}=0$ and $\theta_{0}=\pi$, correspond to the bare states $|1\rangle$ and $|2\rangle$. As $\ap$ and $|\af|$ increase and decrease, respectively, the states originating in $\1$ and $\2$ follow a piecewise spiral trajectory towards the equator of the Bloch sphere. They cross the equator as soon as $\af$ changes sign, and finally interchange with each other when $\ap$ decreases, approaching zero, while $|\af|$ increases. An example of such a piecewise walk on the Bloch sphere is shown in Fig.\ref{FigSpheres}(b).

\begin{figure}[b]
\centering
    \includegraphics[width=0.95\columnwidth]{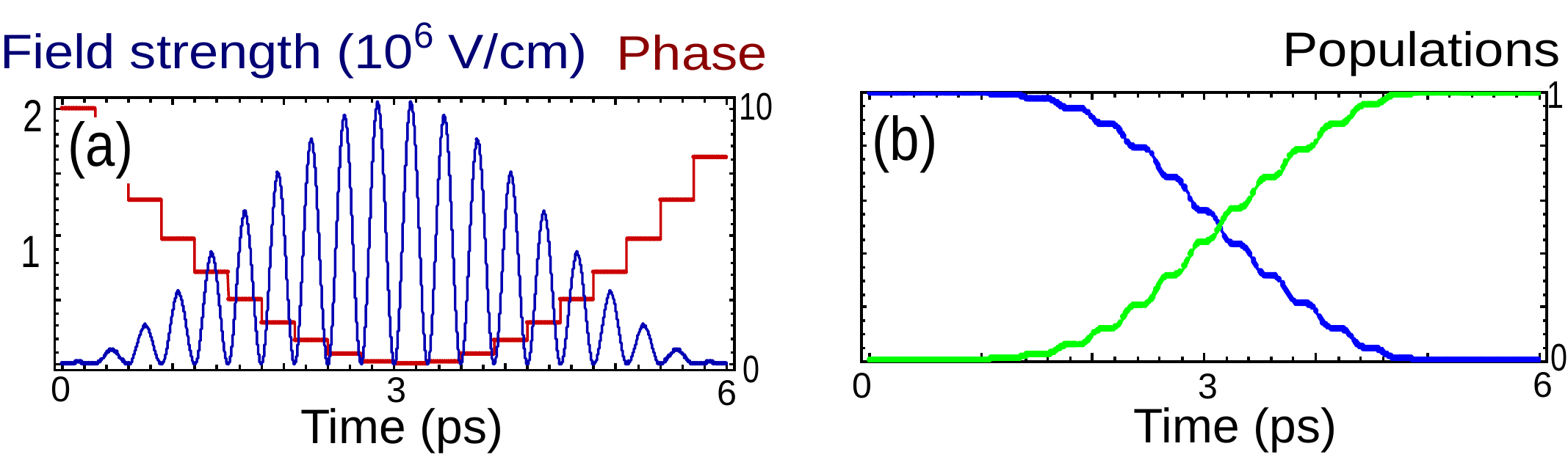}
    \caption{(Color online). (\textbf{a}) The amplitudes (oscillatory, blue)
and phases (piecewise parabola, red) of the driving field.
    (\textbf{b}) The populations of states $\1$ (falling, blue) and $\2$ (rising, green) during the PAP process. }
\label{FigPassages}
\end{figure}

The above analysis shows that the adiabatic following, similar to that implemented with the continuous chirped pulse, can be executed by a sequence of pulses with slowly varying amplitudes, and with the absolute carrier phase changing from pulse to pulse in a non-linear way (i.e. decreasing $\af$ in the first half of the process and increasing $\af$ in the second). An example of such AP, corresponding to the piecewise population transfer in atomic Rb, is shown in Fig.\ref{FigPassages}. Here, the field is given by a sequence of 20 femtosecond pulses. For the $k$-th pulse ($k\in[0,20]$)
\begin{equation}
\label{field-theor}
E_k(t) = A_k \cos\left[\omega_{0} t + \Phi _{k} \right] \times \sin^2\left[\pi\frac{t-t_k}{\tau}\right]
\end{equation}
where $t_k$ marks the beginning of the $k$-th pulse, with $\tau=300$ fs being its full duration. The $A_k$ amplitude represents the train envelope parametrized as a Gaussian of 3 ps width (\textsc{fwhm}). The frequency $\omega_0$ is resonant with the electronic transition between the states $\1 = 5s_{1/2}$ and $\2=5p_{1/2}$ of Rb. The ``piecewise chirp'' of the pulse train is determined by the extra phase factor $\Phi _{k}=\A(k-k_0)^2/2$, where $k_0=11$ and $\A=0.2$. As the pulse sequence proceeds,  the pulse-to-pulse phase change, $\af=\A (k-k_0) + \A/2$, smoothly evolves from large negative values to large positive values. Our simulations show that as long as the conditions of piecewise adiabaticity (\ref{ClassicalAdiabaticity}) are maintained, the population transfer is robust with respect to the pulse shapes, their intensities, and the exact value of the piecewise chirp $\A$.

The population transfer by piecewise chirping described here is a particular example of the Piecewise Adiabatic Passage (PAP) concept introduced in Ref.\cite{Shapiro07}. In the limit of infinitesimal rotations, the present process becomes equivalent to the usual continuous frequency chirping, associated with the quadratic phase change of the carrier phase in time. In this limit, the coarse-grained adiabaticity condition (\ref{ClassicalAdiabaticity}) reduces, up to a numerical factor, to the familiar quantum adiabaticity condition (\ref{QuantumAdiabaticity}).

In the proof-of-principle experiment, we studied the efficiency of the one-photon excitation of atomic rubidium ($^{85}$Rb) from the ground to the first excited electronic state, $5s_{1/2}$ to $5p_{1/2}$, respectively. The effect of a single frequency-chirped laser pulse was compared with the effect of a short pulse train described above. The pulses were produced by a standard Ti:Sapphire regenerative amplifier and had a spectral width of 7.1 nm (\textsc{fwhm}). We tuned the central wavelength of the laser to 795 nm, resonant with the $5s_{1/2} \rightarrow 5p_{1/2}$ transition. To generate a continuous or a piecewise chirp, the original pulse was spectrally shaped using a home made pulse shaper based on a double-mask liquid crystal spatial light modulator in the 4-$f$ configuration\cite{Weiner00}. The excitation beam was focused with  a long focal length lens ($f=$ 100 cm) onto a cloud of rubidium atoms continuously evaporated from the rubidium dispenser in a vacuum chamber equipped with the time-of-flight ion detector.

\begin{figure}
  \includegraphics[width=1.0\columnwidth]{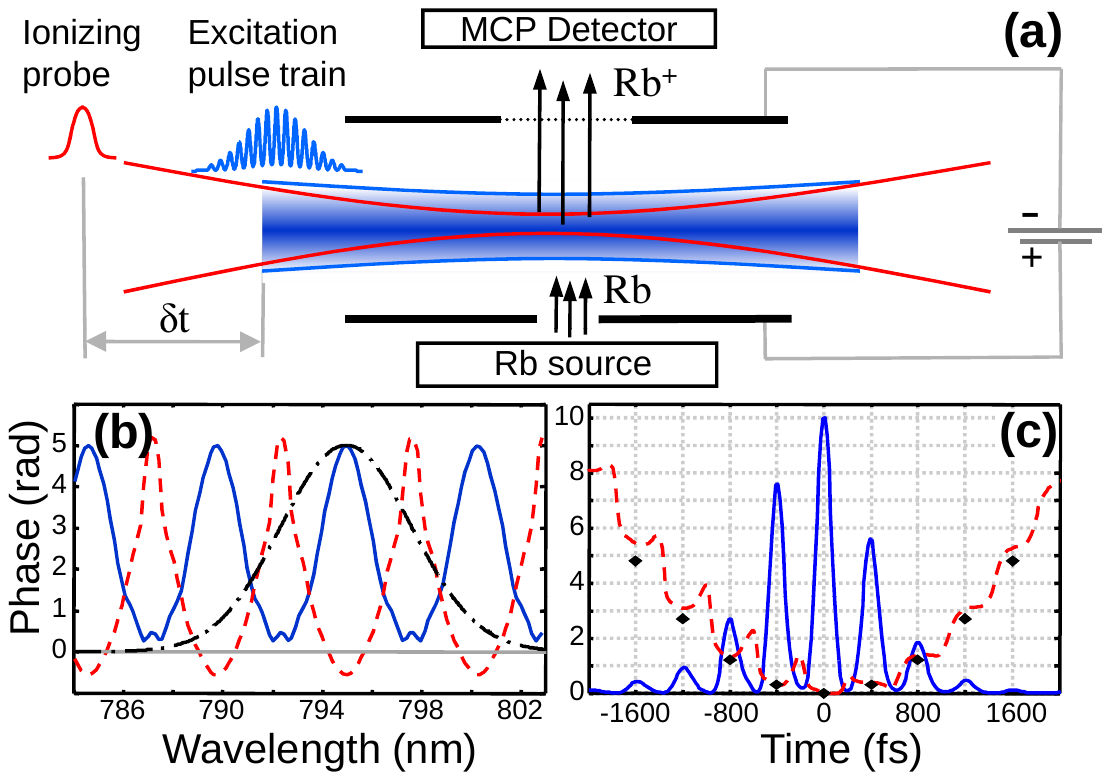}\\
  \caption{(Color online). (\textbf{a}) Scheme of the experimental setup. Rb atoms first interact with a loosely focused pulse train (wide blue) and, after a short time delay ($\delta t=2.5$ ps), are ionized by a tightly focused probe pulse (narrow red). The total amount of $^{85}$Rb$^{+}$ is measured for each excitation pulse with a multi-channel plate detector (MCP); (\textbf{b}) Example of the applied spectral amplitude and phase shaping for generating a train of 9 pulses with a piecewise chirp. Solid (blue) and dashed (red) lines show the amplitude and phase modulation, respectively. Dash-dotted line (black) represents the power spectrum of the original transform-limited pulse; (\textbf{c}) \textsc{frog} retrieval of the temporal field amplitude (solid blue) and phase (dashed red) for a pulse train of 9 pulses after applying the shaping shown in (b) Black dots correspond to the calculated quadratic phase of the pulses, $\Phi _{k}$. All amplitudes are shown in arbitrary units.}
  \label{fig_Setup}
\end{figure}

To determine the population of the excited state, we ionized the atoms by a second, probe pulse tuned to 1300 nm and arriving in the chamber 2.5 ps after the excitation pulse. The ion signal provides good selectivity between the ground and excited state Rb atoms because the ionization of $5s_{1/2}$ requires two more probe photons than that of $5p_{1/2}$. The power of the probe pulse was lowered to less then 0.1 $\mu $J so as to produce no detectable ions from the atoms in the ground state. To ensure ion sampling from the region of uniform excitation field strength, we confined the interaction region in the longitudinal direction by a 3 mm aperture placed between the rubidium dispenser and the laser beam (fig.\ref{fig_Setup}(a)). The suppression of the transverse spatial averaging was done by focusing the probe beam down to a smaller spot size than the size of the 795 nm beam ($1/e^{2}$ beam diameters of 180 and 470 $\mu $m, respectively). Finally, the effect of the laser power fluctuations was eliminated by recording the energy of each excitation pulse together with the corresponding ion count.

We used a home made spectral pulse shaper for introducing continuous frequency chirp and creating the pulse trains required for PAP. The resolution of the pulse shaper of 0.17 nm per pixel allowed us to split the original pulse into a sequence of up to 9 well separated pulses. Continuous chirping was applied by means of the quadratic phase-only modulation, $\phi (\omega )=\alpha (\omega -\omega _{0})^{2}$, where $\alpha $ is the linear chirp and $\omega _{0}$ is the center frequency of the pulse. Generation of a pulse train requires both the amplitude and phase modulation of the field spectrum. A sequence of $N$ replicas of the original transform-limited pulse with the real amplitudes $A_k$ and phases $\omega _{0}t+\Phi _{k}$, separated by the time interval $\tau $ were obtained with the following complex spectral mask:
\begin{equation}\label{Eq_Shaping}
    S(\omega_{n})=\sum_{k} A_{k} e^{-i(\omega_{n} \tau + \Phi_{k})}/|S|,
\end{equation}
where the frequency $\omega _{n}$ corresponds to the $n$-th pixel of the pulse shaper, and the normalization factor $|S|$ was used to satisfy $|S(\omega_{n} )|^{2}\leq 1 \text{\ }\forall \text{\ } n$. Examples of the applied spectral shaping and the corresponding pulse train, characterized by the method of frequency resolved optical gating (\textsc{frog}), are shown in Fig. \ref{fig_Setup}(b,c).

\begin{figure}
  \includegraphics[width=1.0\columnwidth]{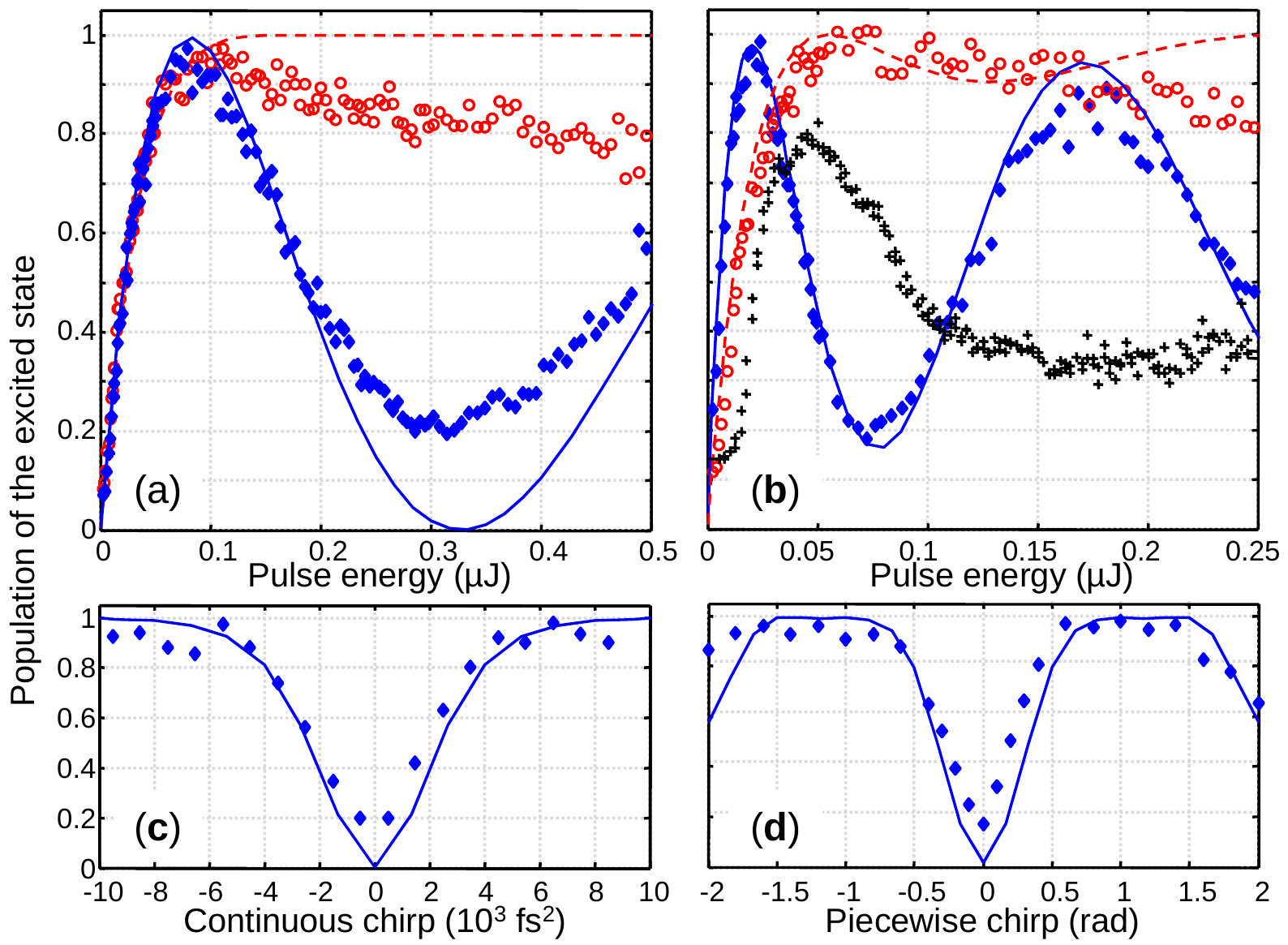}\\
  \caption{(Color online). Measured (dots) and calculated (lines) population of the excited state for a single pulse (\textbf{a,c}) and a train of 7 pulses (\textbf{b,d}). In (\textbf{a}) and (\textbf{b}), the efficiency of the population transfer is shown  as a function of the integrated pulse energy. In both cases, the oscillatory data (blue diamonds and solid lines) reflect the familiar Rabi oscillations, while the second set of data points (red circles and dashed lines) corresponds to the continuously (\textbf{a}) and piecewise (\textbf{b}) chirped excitation with $\alpha =20\times10^{3}$ fs$^{2}$ and $\A=1$ radian, respectively. Black crosses in (\textbf{b}) represent the observed signal for a random (instead of quadratic) distribution of phases $\Phi _{k}$ in the pulse train. In (\textbf{c}) and (\textbf{d}), efficiency of the population transfer is shown as a function of the continuous and piecewise chirp, respectively. In both cases, the energy of the excitation field corresponds to the first minimum of the respective Rabi oscillation. Experimental signal on all 4 plots is normalized to the maximum of Rabi oscillations in plot (a).}
  \label{fig_Results}
\end{figure}

In Fig. \ref{fig_Results} we show the measured ion count corresponding to the population of the $5p_{1/2}$ state of rubidium for various parameters of the excitation field. The oscillatory (blue) curve in plot (a) represents the well known Rabi oscillations as a function of the energy of a single transform-limited pulse. This and all other experimental data points were normalized to the first maximum of this curve at around 0.1 $\mu $J. In plot (b), the similar oscillatory dependence corresponds to the train of 7 pulses with Gaussian envelope of amplitudes and zero piecewise chirp, namely $A_{k}=\exp(-k^{2}\ln{2}/4) \text{\ and\ } \Phi _{k}=0$ where $-3 \leq k \leq 3$. The temporal separation between the pulses in the train was 400 fs. As expected, the integrated pulse area of the train is larger than the corresponding area of a single pulse with the same total energy. The difference stems from the different scaling of the pulse area and pulse energy with the field amplitude (first and second power, respectively), resulting in the smaller period of Rabi oscillations for the piecewise excitation.

When either continuous or piecewise chirp is applied to a single excitation pulse or a series of pulses, respectively, the amount of excited Rb atoms ceases to oscillate with the pulse energy. In both cases it saturates at the maximum value and becomes insensitive to the excitation field strength in the typical AP way (red circles in Fig. \ref{fig_Results}(a,b)). We attribute the decay of both the amplitude of the Rabi oscillations and the saturated signal to the weak pre-pulse generated by our laser system 3 ns prior to the main pulse. To demonstrate the significance of the quadratic phase in the piecewise excitation, in Fig. \ref{fig_Results}(b) we also plot the observed signal for a pulse train with random $\Phi_{k}$ (black crosses). Even though the mutual coherence between the pulses is preserved regardless of the random phase jumps, such pulse sequences do not result in the AP-like saturation of the population of the excited state.

The stability of the adiabatic passage can be further seen in the dependence of the transfer efficiency on the magnitude of the chirp. In both the continuous and piecewise scheme, we set the unchirped pulse area to $2\pi $, which corresponded to the energy of 0.31 and 0.07 $\mu $J, respectively. The result of scanning the conventional chirp from $-10^{4}$ to $+10^{4}$ fs$^{2}$ is similar to scanning the piecewise chirp $\A$ between -1.5 and +1.5 radians, attesting to the similar mechanisms of the two processes. Note the decay of the excited state population at $|\A|>1.5$ rad in plot (d), caused by the breakdown of the piecewise adiabaticity due to the increasingly high phase increments from pulse to pulse in the pulse train. The results of numerical simulations are shown along with the corresponding experimental results as solid lines in Fig. \ref{fig_Results}. The only single fitting parameter for all calculations was the area of the excitation beam. \textsc{frog} traces were used to define the temporal profile of the pulses.

In contrast to adiabatic transfer with continuously chirped pulses, PAP can be easily generalized to a wide class of cases. For example, one can consider replacing states $\1$ and $\2$ by wave packets composed of many individual eigenstates. Piecewise excitation of such systems could be advantageous because, by their very nature, wave packets often evolve periodically in time. Viewed from the spectral perspective, this extension of PAP is related to the ''Coherently Controlled Adiabatic Passage'' schemes, proposed earlier for controlling population transfer between wave packets\cite{Kral07}. We have simulated numerically a number of situations in which states $\1$ and $\2$ correspond to an isolated vibrational level and a wave packet composed of several vibrational levels, respectively. The population transfer is executed with a train of femtosecond pulses separated in time by the wave packet's vibrational period. Quadratically varying the carrier phase of the pulses enables robust and complete population transfer into the excited state wave packet, whose shape can be controlled by spectrally shaping the individual pulses in the train. The availability of such robust and selective population transfer in multilevel systems positions PAP as a powerful tool in controlling molecular dynamics.


\begin{acknowledgements}
The authors would like to thank Avi Pe'er and Jun Ye for helpful discussions, and Qun Zhang for help with the experimental setup. This work has been supported by the CFI, BCKDF and NSERC.
\end{acknowledgements}


\end{document}